# Using Artificial Intelligence to Augment Science Prioritization for Astro2020

Brian Thomas[1], Harley Thronson[2], Andrew Adrian[1], Alison Lowndes[3], James Mason[2], Nargess Memarsadeghi[2], Shahin Samadi[4], and Giulio Varsi[5]

1. NASA Headquarters, 300 E Street SW, Washington DC
2. NASA Goddard Space Flight Center, 8800 Greenbelt Rd, Greenbelt, MD
3. NVIDIA, Green Park, Reading, United Kingdom
4. INNOVIM LLC, Greenbelt, MD
5. NASA Headquarters, retired

Corresponding author: Brian.A.Thomas@nasa.gov

*Submitted in response to the State of the Profession APC call;*
*Decadal Survey of Astronomy and Astrophysics, 2020*

Attn: Enabling Foundation for Research

# Executive Summary

Science funding agencies (NASA, DOE, and NSF), the science community, and the US taxpayer have all benefited enormously from the several-decade series of National Academies' Decadal Surveys. These Surveys are one of the primary means whereby these agencies may align multi-year strategic priorities and funding to guide the scientific community. They comprise highly regarded subject matter experts whose goal is to develop a set of science and program priorities that are recommended for major investments in the subsequent 10+ years. They do this using both their own professional knowledge and by synthesizing details from many thousands of existing and solicited documents.

Congress, the relevant funding agencies, and the scientific community have placed great respect and value on these recommendations. Consequently, any significant changes to the process of determining these recommendations should be scrutinized carefully. That said, we believe that there is currently sufficient justification for the National Academies to consider some changes. We advocate that they supplement the established survey process with predictions of promising science priorities identified by application of current Artificial Intelligence (AI) techniques These techniques are being applied elsewhere in long-range planning and prioritization.

We present a proposal to apply AI to aid the Decadal Survey panel in prioritizing science objectives. We emphasize that while AI can assist a mass review of papers, the decision-making remains with humans. In our paper below we summarize the case for using AI in this manner and suggest small inexpensive demonstration trials, including an AI/ML assessment of the white papers submitted to Astro2020 and backcasting to evaluate AI in making predictions for the 2010 Decadal Survey.

*Keywords* : Artificial Intelligence, Machine Learning, Pattern Recognition, Strategic Planning, Decadal Survey, Astro2020

**CONTENTS**

I. **Introduction and Motivation**

Astro2020, the current National Academies' Astronomy and Astrophysics Decadal Survey, is the sixth in the series of generally similar long-range strategic prioritization activities intended to help the funding agencies identify the highest priority goals in astronomy for the subsequent 10+ years. The Surveys have also highlighted notional mission concepts to achieve these goals, assessed the health of the profession, prioritized technology areas for investment, and judged the relative balance in the portfolio of small, medium, and large missions.

The Decadal Surveys have become the "gold standard" for strategic planning in the sciences (and elsewhere) largely because of a handful of highly respected characteristics: (1) significant, multi-year commitment to their creation by the Academies, the funding agencies, and the individual participants; (2) participation of highly regarded subject matter experts chosen to represent key demographics and professional specialties; (3) transparency and openness to community input, including mid-decade reviews; and (4) a record of success, as shown by the national commitment to fulfill the recommendations of previous Surveys.

Although widely and highly regarded, we believe that there is room for improvement in the process of developing the Survey. The principal challenge stems from the panelists' need to assess a large and ever-growing amount of relevant information. Materials have increased in both variety and number. The relevant information to support decisions of the Survey cannot be uncovered based solely on a panelist's experience and by reading the key papers published in a few premier journals. A prudent panel will also consider papers published in all relevant journals, important blogs, websites, key unpublished research papers (such as may be found on ArXiv.org), organizational whitepapers, documentation about various missions and their success statistics, open source codebases and related statistics, materials detailing use and development of very large archives and so forth. In each of these categories the volume and complexity of the information to consider is growing yearly. Bornmann and Mutz (2015) have estimated the global rate of increase in scientific publication at about 8 to 9% yearly which means the number of research papers published in a given year doubles approximately every 8 to 9 years. Furthermore, this rate of increase has itself been increasing since the end of the 1940s when the yearly rate of increase in publication was significantly lower (~3% per year).

Because of this explosion in relevant data, it is becoming more difficult to know important details and synthesize critical information. Without aid, the panelists no longer have sufficient knowledge of the field. As the pace of research and avenues for publishing relevant data both increase, the problem becomes more acute as new research lies hidden within the haystack of materials.

The Survey process is at a critical juncture: the weight of information must be better managed and existing methods of panel work should be augmented. We believe that artificial intelligence (AI) offers a possible solution to the current challenges and is sufficiently mature to be considered seriously as a new tool for the Survey.

Advances in AI software over the past decade have been impressive. The rapid adoption of AI in recent years often masks the reality of the field: the origin of these technologies was in the late 1950s. While it is true that more powerful AI algorithms and techniques have been developed over these past decades, the advent of readily available processing power and large volumes of usable data that has also made application of AI now practical and useful. Cloud and GPU technologies have provided the needed processing power. Development of techniques to change large unstructured digital documents into a form that may be utilized in training has also been important. In addition, the relatively recent availability of digitized research papers makes application of AI much easier.

Rapid advances in AI are capturing scientific, economic, and public interest (The National Academies 2018). In the realm of big data, machine learning (ML, a branch of AI) has been used to keep up with the ever-growing amount of information. An area of significant promise is the use of these techniques to tackle problems that involve large amounts of unstructured data (e.g., text documents). Natural Language Processing (NLP) techniques have matured and can be applied to document understanding which can more quickly summarize content and reveal connections between documents. Consider the common activity of peer review. Most researchers have good reason to grumble about bias, although now peer review by AI is promising to improve the process, boost the quality of published papers — and save reviewers' time. A handful of academic publishers are piloting AI tools to do everything from selecting reviewers to checking statistics and summarizing a paper's findings. It is important to note that automated software can help review papers, but the decision-making stays with humans (Heaven 2018). HIL (human-in-the-loop) is vital for acceptance of AI as well as success of AI in deployment. In critical-control deployments it is the only way AI can be used, as is the case currently with self-driving car.

These same approaches can be applied to improve the Decadal Survey. The desire to predict discoveries – to have some idea, in advance, of what will be discovered, by whom, when, and where – has been a priority of strategic planning for decades (Clauset, Larremore, and Sintara 2017). Progress has been made and now AI-identified strategic goals are becoming regular augmentations to conventional planning processes. For example, Krenn et al. (2019) demonstrate a method to build a semantic network from published scientific literature, which is used to predict future trends in research and to inspire new, personalized and surprising seeds of ideas in quantum physics.

## II. The Potential for Application of AI to Decadal Surveys

We believe now is the time for AI augmentation to be considered for the Decadal Surveys. For each major scientific research area, the National Academies offers only a single opportunity for major input every decade. Thus, we feel that for the astrophysics Decadal Survey, the scientific strategic planning effort that establishes the norm, the time to consider this capability is now. AI has been demonstrated in many cases to be superior to unaugmented human capabilities in finding patterns and correlations among items such as object/image recognition (Sabour, Frosst ad Hinton 2017; Hinton, Sabour and Frosst 2017), and predictions from large amounts of data (Gebru et al. 2017).

Generally, the algorithms do not ascertain whether the correlations discovered are symptoms of causality or randomness. Hence, we propose to use the algorithms to discover hidden correlations and provide suggestions to the Survey panelists. The panelists may then assess possible causality and predict which of many possible future endeavors (aka, "missions") has the highest probability of success.

In the context of Astro2020, AI brings a major advantage: it can vastly increase the number of sources of information (inputs) while providing only a few highly condensed materials (outputs) for review by the Survey's panels. Many of these notional inputs may be entirely novel for the Survey, which will increase the breadth and diversity of factors under consideration by AI for making strategic recommendations. AI can also augment the existing inputs, for example, by providing narrative summaries and statistics on the contents of the numerous white papers. Similar efforts are underway in other disciplines, e.g., the SemNet project for Quantum Physics literature (Krenn et al. 2019). Subsections below detail the potential inputs and outputs, respectively.

A. Inputs

The types of inputs for AI algorithms can be surprisingly broad. Text, numbers (e.g., fluxes), images, websites - especially those with Application Programming Interfaces (APIs) for data scraping - and all manner of traditional datasets.

*Astronomy Literature*
The first and most traditional input would be the refereed astronomy literature. This can be accessed easily through the Astrophysics Data System (ADS, http://www.adsabs.harvard.edu). The Natural Language Processing (NLP) tools of AI could be applied to read the papers themselves. Many journals support machine readable tables that could be easily ingested along with the text.

Additional literature could be considered. For example, non-refereed articles are also easily accessed with the ADS API and are flagged as such, making filtering easy. Articles on the history of astronomy, such as biographies and mission histories, may also be valuable. Abstracts for posters and presentations likely do not contain as many results, although certainly are an indicator of research activity. Finally, an especially crucial input would be previous strategic plans: the prior Decadal Surveys and other similar plans developed by NASA, NSF, and DOE. It will then be possible to compare the recommendations with the outcomes, in a similar manner to the mid-term surveys, although using the power of AI augmentation.

*Mission Statistics*
The agencies fund space- and ground-based observatories on small and large scales. On large scales, these selections are often in direct accordance with the Decadal Surveys, but even on the small scales, the targets and research focus areas are often influenced by the Decadal Surveys. Thus, an important input is the impact of past missions. "Impact" is a deliberately broad term that allows for many metrics to play a part. For example, for each mission, data could be input for the number of associated papers, citation count per year, presentation/poster count per year, number of data downloads, number of associated prizes, and public engagement as measured by the number of news articles and/or tweets per year. Some of these data will be relatively easy to obtain while others may be scraped from the web or as a last and likely prohibitively time-consuming - resort, obtained by private communications with mission investigators.

*Code Repositories*
Historically overlooked, open source code repositories contain a wealth of information about what work is being done in astronomy. Many of these efforts dovetail into refereed journal articles, although such articles tend to focus more on results than on methods. Code is the most concise and complete description of a method to obtain a result. There are many metrics of the activity and usefulness of a code repository contained in its metadata. For example, major code hosting platforms such as GitHub, GitLab, and BitBucket track the number of code commits as a function of time, the number of forks, the number of users contributing, the number of reported issues and who reported them, the number of visitors as a function of time, and what other repositories any particular one is dependent upon. There is a growing push in the community to give career credit to researchers who sacrifice time building tools that the astronomy community uses, time that could otherwise be spent publishing papers to get traditional career credit. See, for example, the recommendations in the National Academies 2018 Report entitled "Best Practices for a Future Open Code Policy for NASA Space Science" (Gentemann and Parsons, 2018). Including code repository metrics as input to the Decadal Survey via AI tools is one such way to ensure that these contributions are recognized, encouraged, and supported more fully in the future as well as providing a measure of the impact of particular tools in astronomy.

*Survey White Papers & Costs*
Use of submitted Survey activities and projects whitepapers allows a direct comparison of impact among possible Survey recommendations (or components thereof). For example, a Discovery-class satellite and a sounding rocket experiment will have different scientific impacts and costs. The former will always dwarf the latter in both categories. We may use AI to cluster similar projects and activities together and to find the extent that these projects and activities are in alignment with avowed need in the literature (above). What is important is to:

1. assess the *impact per dollar* wherever possible across many similar projects and activities

   and

2. determine how well these projects and activities' mission goals align with perceived needs in the literature and within the corpus of white papers themselves.

We should not fall into the trap of "ranking missions," as Forbes might be ranking billionaires. The focus here should be on identifying the most productive/compelling areas of scientific/astronomical investigation for the upcoming 10 years (*e.g.*, X-rays or gravitational waves or gamma-rays or ……). Later the best approach (*aka* "mission") can be devised by using standard, old fashioned "system engineering."

### B. Outputs

Each of these outputs is designed to be an augmentation to the existing Survey process: covering blind spots, providing insights, and suggesting priorities for consideration by Survey personnel.

*Narrative Summaries*
Natural Language Processing (NLP) tools have already demonstrated some truly surprising capabilities. For example, the Beta Writer project ingested the entire library of literature on lithium ion batteries and created an entirely machine-produced, human-readable book summarizing the topic (Beta Writer 2019). This same technology could be used to ingest the entire library of astronomy literature, or any particular subselection - for example, only the past decade - and produce a brief summary with tunable length. In fact, as an interesting experiment, the same could be done with the numerous white papers submitted for Astro2020.

*Citation Statistics*
Much simpler, and only requiring traditional data analytics, the citation data could be compressed into simple metrics. These citations include cross-references in refereed articles and code repository dependencies. Rather than producing an h-index, which is tied to an individual author, an equivalent could be done for research topic areas, missions, and programmatic methods. Those that rate highly would be worth investigating as avenues of future strategic investment.

*Identification of High-Impact Topic Areas*
These would not be normalized by cost. Instead, it would be used as a sanity check. The results should verify our expectations. For example, Hubble Space Telescope, multi-messenger astronomy, and exoplanets had large impacts across all metrics.

*Identification of High-Impact-per-Dollar Topic Areas*
These would ultimately be suggested priorities for consideration. It is likely that many of the identified topics or instruments would be a verification of what the panelists already knew, which would serve to bolster confidence given the independence of the method. However, some of the identifications may come as a surprise and warrant further investigation by the panelists. If born out, these are suggestions that may not have otherwise found a place in Astro2020.

## III. Proposed Small-Scale Demonstration: Backcasting to the Astro2010 Decadal Survey

An obvious, affordable, and instructive opportunity to demonstrate the value, and challenges, of applying AI capabilities to Decadal Survey strategic prioritization would be to backcast to the last Decadal Survey in 2010.

As part of our proof-of-concept assessments for this white paper, Figure 1 below shows a preliminary example of one type of graphic output of applying AI to a small trial data set. A clustering has been combined with semantic extraction of common theme(s) of papers. Without human labelling it has auto-identified clusters of papers in six astronomy topic areas. Each dot represents a published paper and strong connections between papers are shown as lines.

We made no effort to choose representative research papers for the input, rather we were assessing the process, available software, and graphics displays, and learning to tweak the AI "rules." Nevertheless, there is some interesting clustering and apparent relationships among disparate topic areas. It is easy to see the potential in an approach such as this. We can use the results to cross-validate choices in a prior Decadal Survey by comparing the suggested topics from review articles, and associated materials, published prior to that given Survey.

For a subsequent, more ambitious attempt, the AI software would be trained on a more complete sample of published articles: review papers published from 2000 to 2010, for example. Furthermore, future attempts would include other input information (blogs, news articles, primary publications) and we would apply and test various filtering criteria to the content utilized for training and optimization.

We would also apply a more advanced and scalable solution than that of our test case shown in Figure 1. For example, we can apply a combination of clustering and XGBoost (gradient boosted decision trees) with the latest processing units and libraries such as RAPIDS (https://rapids.ai) and GPU-accelerated BERT (a language representation model of unlabeled text). Widely used and accepted by the NLP community[1], BERT is proven to achieve impressive levels of predictive accuracy as it can analyze entire sentences to more accurately learn the context of a given word based on its surroundings in both directions. Further improved techniques include XLNet from Google & Carnegie Mellon University (Yang et al. 2019). Other possibilities include using word2vec software (https://github.com/tmikolov/word2vec) for gaining insight and capturing knowledge among research papers (Tshitoyan et al. 2019).

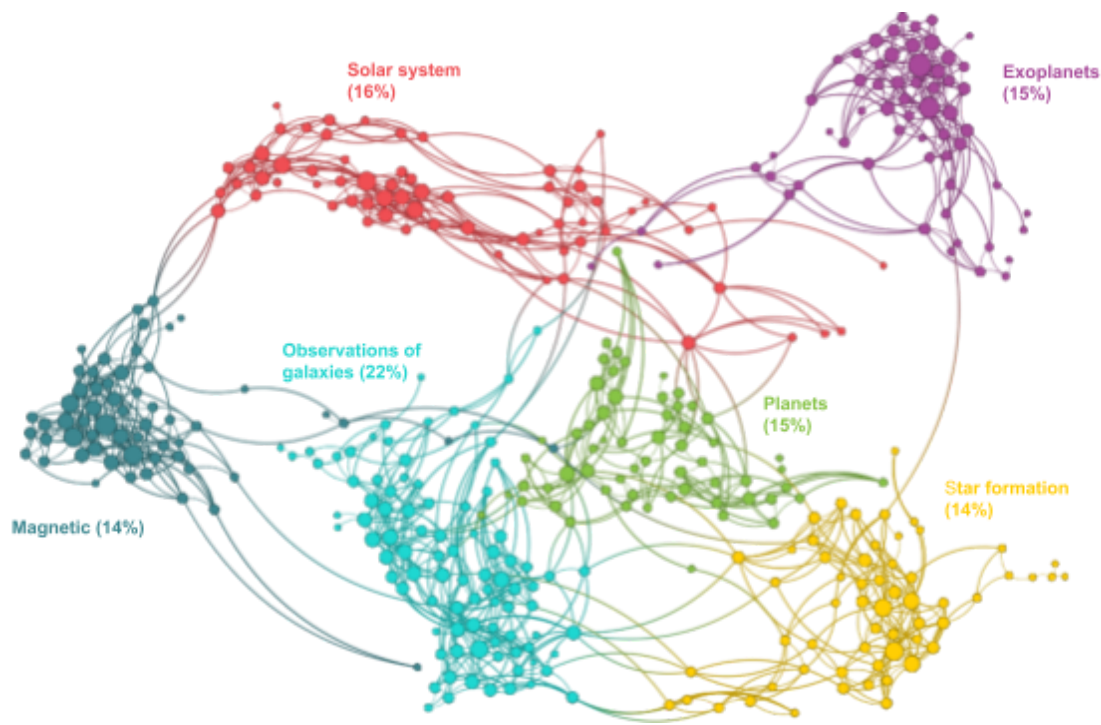

*Figure 1. Example ML-based clustering exercise on a small sample of astronomical papers. Six topic areas of significance were indicated. A similar example of what is possible may be found in the Paperscape project[2].*

As our eventual goal is to predict future priorities and trends with AI we will rely on published work created by humans. While this data is essential to training the AI system we are aware that there are pitfalls to be avoided, as is true with any AI method. We must configure any such AI system wisely by thoughtful examination of our choice in training data, filtering and feature criteria. Sampled inputs can of course incorporate bias which must be carefully assessed. Careful selection along with clear communication of the choices of how the AI is trained can help to alleviate these concerns.

---

[1] see https://towardsdatascience.com/bert-explained-state-of-the-art-language-model-for-nlp-f8b21a9b6270
[2] http://paperscape.org

Sample data we can utilize from astrophysics could cover an extraordinary range of topics from the birth of the Universe to stellar evolution. To begin this initial experiment, we suggest limiting the topic to, for example, galaxy evolution and structure and selection of articles which summarize the state of this field. The goal of this simple trial will be to predict the science priorities for, in this case, "galaxy evolution and structure" in a simulation of the 2010 Decadal Survey and compare with that identified in the actual Survey. This comparison will likely lead to iterations on the adopted training documents, although it would be interesting if our modeled Survey identified a priority that was not in the actual 2010 Survey. That said, it is essential to note that one of the most important products of Decadal Surveys is prioritization among multiple major astrophysics research areas; further improvement of our trial should likewise include multiple topic areas.

IV.     **Summary**

We believe the time is right for The Academies to consider augmenting their process for determining the highest strategic science priorities in Decadal Surveys. The primary challenge is for the panelists to understand the increasingly large number and variety of knowledge they are expected to understand. Happily, we believe that the resources now exist to build a sort of artificial intelligence 'bionic arm' to support their attempt. AI algorithms maturity, GPU and Cloud technologies, NLP techniques, and access to critical input information have all advanced to a point to make such a project feasible for the first time.

We believe that it is well worth the attention of The Academies to consider the use of AI now and that the cost and risk of such an effort is low. We recommend a series of trial demonstrations of AI-augmented software to compare and test various algorithms effectiveness and choices of input information (Section II) to predict future major science goals. Test data already exist for this effort in the form of prior Decadal Survey recommendations, suggested projects and activities, and past published review papers. This effort is to be carried out under the auspices of The Academies and supported by the funding agencies. Software tools vetted by debate, discussion, and trial can, as they are ready, be deployed for subsequent Surveys.

**REFERENCES**

- Beta Writer, 2019, "Lithium-Ion Batteries. A Machine-Generated Summary of Current Research", url: https://doi.org/10.1007/978-3-030-16800-1

- Bornmann, L. and Mutz, R., 2015, Growth rates of modern science: A bibliometric analysis based on the number of publications and cited references. J Assn Inf Sci Tec, 66: 2215-2222. doi:10.1002/asi.23329
- Clauset, A., Larremore, D.B., and Sintara, R., 2017, "Data-driven predictions in the science of science", Science, vol. **355**, p. 477–480
- Heaven, D., 2018, "The age of AI peer reviews" Nature **563**, 609–610
- Hinton, G., Sabour, S., and Frosst, N., 2018, "Matrix capsules with EM routing", In proceedings of International Conference on Learning Representations, url: https://openreview.net/forum?id=HJWLfGWRb
- Gebru et al., 2017, "Using Deep Learning and Google Street View to Estimate the Demographic Makeup of the US", arXiv:1702.06683 [cs.CV]
- Gentemann and Parsons, 2018, "Best Practices for a Future Open Code Policy for NASA Space Science", url: https://sites.nationalacademies.org/SSB/CurrentProjects/SSB_178892
- Krenn et al., 2019, "Predicting Research Trends with Semantic and Neural Networks with an application in Quantum Physics" arXiv:1906.06843v1 [cs.DL]
- National Academies, 2018, "The Frontiers of Machine Learning: 2017 Raymond and Beverly Sackler U.S.–U.K. Scientific Forum", doi:10.17226/25021
- Sabour, S., Frosst, N., and Hinton, G., 2017, "Dynamic Routing Between Capsules", arXiv:1710.09829 [cs.CV]
- Tshitoyan, V. et al., 2019, "Unsupervised word embeddings capture latent knowledge from materials science literature", *Nature* 571, 95–98
- Yang, Z. et al., 2019, "XLNet: Generalized Autoregressive Pretraining for Language Understanding", arXiv:1906.08237 [cs.CL]